\documentclass{emulateapj}
\usepackage{psfig}

\newcommand{\bm}[1]{\mbox{\boldmath{$#1$}}}
\newcommand{\be}{  \begin{eqnarray} }
\newcommand{\ee}{  \end{eqnarray}} 
\newcommand{\bd}{  \begin{displaymath} }
\newcommand{\ed}{  \end{displaymath} }
\newcommand{\emf}{\mathcal{E}}

\newcommand{\bfourx}{\tilde{B}_x({\bm k})}
\newcommand{\bconjx}{\tilde{B}_x^{\ast}({\bm k})}
\newcommand{\bfoury}{\tilde{B}_y({\bm k})}
\newcommand{\bconjy}{\tilde{B}_y^{\ast}({\bm k})}

\begin{document}
\title{Sustained Magnetorotational Turbulence in Local Simulations of Stratified Disks with Zero Net Magnetic Flux}
\author{Shane W. Davis\altaffilmark{1,2}, James M. Stone\altaffilmark{3}, and
Martin E. Pessah\altaffilmark{1}}
\altaffiltext{1}{Institute for Advanced Study,
Einstein Drive, Princeton, NJ 08540}
\altaffiltext{2}{Chandra Fellow}
\altaffiltext{3}{Department of Astrophysical Sciences, Princeton University, Princeton, NJ 08544, USA}

\begin{abstract}

  We examine the effects of density stratification on
  magnetohydrodynamic turbulence driven by the magnetorotational
  instability in local simulations that adopt the shearing box
  approximation.  Our primary result is that, even in the absence of
  explicit dissipation, the addition of vertical gravity leads to
  convergence in the turbulent energy densities and stresses as the
  resolution increases, contrary to results for zero net flux,
  unstratified boxes.  The ratio of total stress to midplane pressure
  has a mean of $\sim0.01$, although there can be significant
  fluctuations on long ($\gtrsim 50$ orbit) timescales.  We find that
  the time averaged stresses are largely insensitive to both the
  radial or vertical aspect ratio of our simulation domain.  For
  simulations with explicit dissipation, we find that stratification
  extends the range of Reynolds and magnetic Prandtl numbers for which
  turbulence is sustained.  Confirming the results of previous
  studies, we find oscillations in the large scale toroidal field with
  periods of $\sim 10$ orbits and describe the dynamo process that
  underlies these cycles.

\end{abstract}

\section{Introduction}
\label{intro}

The magnetorotational instability (MRI) plays an important role
in determining the angular momentum transport rate (and therefore
accretion rate) in most astrophysical disks \citep{bh98}.
Therefore it is of considerable interest to understand what
determines the saturation amplitude of the MRI in the nonlinear regime.
Investigations of this question generally utilize numerical methods to
study the time-dependent MHD in the local, shearing box approximation.

From the first three-dimensional studies \citep{hgb95}, it has been
known that for uniform density the saturation amplitude depends on
parameters such as the
net flux and geometry of the magnetic field threading the domain
\citep{san04}.  More recently, there has been considerable interest in
the effect of microscopic dissipation, such as Ohmic resistivity and
Navier-Stokes viscosity, on the saturation amplitude with various
initial field geometries \citep{fro07,ll07,sh09}, as well as the effect of
the radial extent of the domain \citep{bod08,jyk09,gua09}.

One particularly important and puzzling result is that in the special
case of no net magnetic flux with no explicit dissipation, the
saturation amplitude of the MRI decreases with increasing resolution
\citep{fp07,pcp07}.  In this case, it appears the amplitude of
the microscopic diffusivities determines the saturation amplitude
of the MRI.  Although this result is of considerable interest from a
theoretical perspective in understanding the MRI and MHD turbulence,
it is not yet obvious it has application to real disks, in which the
magnetic flux is unlikely to be zero in every local patch for all time
(Sorathia et al. 2009), and which are vertically stratified.

Although the saturation of the MRI has been studied in the local
shearing box approximation in vertically stratified disks
\citep{bra95,sto96}, these early studies lacked sufficient
computational resources to perform a systematic convergence study, or
evolve the disk for hundreds of orbital times in order to measure
accurately the saturation amplitude.  In this paper, we use modern
numerical methods to revisit the saturation of the MRI in vertically
stratified disks\footnote{Throughout the paper we describe our
  simulations as stratified, even though we assume an isothermal
  equation of state.  In this text stratification simply refers
  to the density stratification which is the result of
  vertical gravity in our equations. It is not a reference to
  the entropy gradient.} with no initial net magnetic
flux.  Interestingly, in this case we find quite different results
compared to the unstratified boxes studied by \citet{fp07}.  In the
stratified boxes studied here, the stress converges with numerical
resolution even with no explicit dissipation, In fact, with explicit
dissipation, we find in stratified disks there can be significant and
sustained turbulence at magnetic Reynolds numbers that suppress the
turbulence in unstratifed disks \citep{fro07}.  These results seem to
be a consequence of an MHD dynamo that operates in stratified disks,
and we explore the properties of this dynamo in this paper.

This paper is organized as follows.  In \S\ref{meth} we summarize the
most relevant properties of our numerical methods and describe our
Fourier analysis.  In \S\ref{res} we report our results: the outcome
of our resolution study in \S\ref{std}; the dependence on the vertical
and radial aspect ratios in \S\ref{tall} and \S\ref{wide}; and the
effects of adding finite dissipation in \S\ref{dissip}.  In
\S\ref{disc} we discuss the nature of the dynamo driving the
sustained turbulence, and in \S\ref{conc} we summarize our
conclusions.

\section{Method}
\label{meth}

We use Athena \citep{gs05,gs08,sto08} for all calculations presented
in this work.  We perform 3d MHD simulations, adopting the local
shearing box formalism and including vertical gravity.  We refer the
reader to Stone \& Gardiner (2009) for a detailed
discussion of the equations, algorithms, and boundary conditions
specific to the shearing box, as well as a description of their
implementation in Athena.  Here we just summarize the basic equations
and the most relevant features for our current work.

The local shearing box approximation adopts a frame of reference
located at a fiducial radius corotating with the disk at orbital
frequency $\Omega$.  In this frame, the equations of ideal MHD are
written in a Cartesian coordinate system $(x,y,z)$ that has unit
vectors $({\hat{\bm i}}, {\hat{\bm j}}, {\hat{\bm k}})$ as
\begin{eqnarray}
\frac{\partial \rho}{\partial t} +
{\bm\nabla\cdot} [\rho{\bm v}] & = & 0,
\label{eq:cons_mass} \\
\frac{\partial \rho {\bm v}}{\partial t} +
{\bm\nabla\cdot} \left[\rho{\bm vv} + {\sf T}\right] & = &
\rho \Omega^{2}(2qx{\hat{\bm i}} - z{\hat{\bm k}}) - 2\Omega {\hat{\bm k}} \times \rho {\bm v},
\label{eq:cons_momentum} \\
\frac{\partial {\bm B}}{\partial t} -
{\bm\nabla} \times \left({\bm v} \times {\bm B}\right) & = & 0,
\label{eq:induction}
\end{eqnarray}
where ${\sf T}$ is the total stress tensor
\begin{equation}
  {\sf T} = (p + B^{2}/2){\sf I} - {\bm B^{\rm T}}{\bm B},
\label{eq:stresstensor}
\end{equation}
${\sf I}$ is the identity matrix, $\rho$ is the gas density, $p$
is the gas pressure, ${\bm B}$ is the magnetic field, ${\bm v}$ is the
velocity and $B^{2} = {\bm B} \cdot {\bm B}$.  The shear parameter $q$
is defined as
\begin{equation}
 q = - \frac{d {\rm ln} \Omega}{d {\rm ln} r}
\label{eq:shearparam}
\end{equation}
so that for Keplerian flow $q=3/2$.  We adopt an isothermal equation
of state with $p=c_{\rm s}^2 \rho$.  

In \S\ref{dissip} we also present simulations which include terms for
constant scalar viscosity and resistivity.  The viscous term is ${\bm \nabla
  \cdot} {\sf M}$ with
\be
{\sf M}_{ij} = \rho \nu \left(\frac{\partial v_i}{\partial x_j}+
\frac{\partial v_j}{\partial x_i} - \frac{2}{3}\delta_{ij}
{\bm \nabla\cdot v} \right),
\ee
and the resistive term is $-{\bm \nabla} \times (\eta {\bm
  \nabla}\times{\bm B})$ when added to the right-hand side of
(\ref{eq:cons_momentum}) and (\ref{eq:induction}), respectively.

These sets of equations admit the well know solution corresponding to
(linearized) uniform orbital motion
\be
{\bm v}_{K} = -q\Omega x {\hat{\bm j}},
\ee 
which is used for the initial condition.  The initial equilibrium
density configuration is Gaussian with
\be
\rho=\rho_0 \exp\left(-\frac{z^2}{H^2}\right),
\ee
where $H=\sqrt{2} c_s/\Omega$ is the scale height of the disk. For
consistency with previous work \citep{sto96}, we take
$\Omega=10^{-3}$, $c_s=5 \times 10^{-7}$, and $\rho_0=1$, yielding
$p_0=5 \times 10^{-7}$ and $H=1$.  All simulations are initialized to
have a weak magnetic field with a ratio of midplane gas pressure to
magnetic pressure $\beta=2 P_0/B^2 =100$. The configuration is a
vertical field with zero net magnetic flux that varies sinusoidally
in the radial direction.

We adopt boundary conditions which are shearing periodic in $x$, and
periodic in both $y$ and $z$.  Clearly, the periodic assumption in $z$
is physically unrealistic in a stratified box.  Of course, one
advantage of this assumption is computational expediency, but the main
motivation is to give us some level of `control' over the evolution of
magnetic flux in the simulation domain.  Vertically periodic boundary
conditions are useful in that the mean (volume averaged) toroidal
field is conserved (i.e. remains zero).  Outflow boundary conditions
will necessarily introduce electromotive forces (EMFs)
at the vertical boundary which can
drive growth of non-zero $\langle B_y\rangle$.  Such mean field
evolution is plausible in real disks, but we worry about spurious
growth in $\langle B_y\rangle$ due to the manner in which outflow
boundary conditions are implemented.  
These considerations are important because
the presence of mean azimuthal field may enhance or sustain
turbulence \citep{hgb95}.  Since one of our primary goals is to examine
the robustness of MRI turbulence in stratified disks, this
prescription, which prevents the grow of a (box integrated) mean
field, represents a conservative approach.

All simulations make use of Athena's orbital advection scheme (Stone
\& Gardiner, 2009), allowing us to consider domains with large
radial extent.  Orbital advection schemes \citep{mas00,jgg08} take
advantage of the fact that Equations
(\ref{eq:cons_mass}-\ref{eq:induction}) can be split into two systems,
one of which corresponds to linear advection operator with velocity
${\bm v}_K$ and another with only involves the fluctuations $\delta
{\bm v}={\bm v}-{\bm v}_K$. The integration of linear advection
operator is very simple and not subject to a Courant-Friedrich-Lewy
(CFL) condition.  Since $\delta {\bm v} \ll {\bm v}_K$ near the
boundaries, the CFL condition in the second system is much less
restrictive than in standard algorithms, particularly for domains with
large radial extent.  It also has the advantage of removing the
systematic variation of truncation error introduced by the shear,
which can lead to spurious effects \citep{jgg08}.

\subsection{Fourier Analysis}
\label{fft}

We utilize a number of diagnostic tools to analyze the simulation
output, including Fourier analysis.  This is straightforward in a
periodic domain, but in a shearing periodic system, the basis
functions are shearing waves with a time dependent wavevector.  This
complication can be handled with simple remappings before and after
Fourier transforming, as outlined in \citet{hgb95}.

The quantities of principal interest will be the power density spectra
(PSDs) of the magnetic and kinetic energies.  Although the PSD is
highly anisotropic in $k$-space, we still find it useful to plot
shell averaged Fourier amplitudes.  For example, we define the shell 
averaged power spectrum of the  magnetic field  as 
\be
B^2_k \equiv 4\pi k^2 |\tilde{B}(k)|^2,
\ee
where $|\tilde{B}(k)|^2$ denotes the average of $|\tilde{B}({\bm
k})|^2$ over spherical shells, and $\tilde{B}({\bm k}) = \int B({\bm
x}) \exp{(-i {\bm k} \cdot {\bm x})} d^3 {\bm x}$ is the Fourier
transform of $B$.  \footnote{Of course, all Fourier analysis in this work
refers to discrete Fourier transformations of discretized data. However,
for the ease of readability, we will use continuous notation throughout
the text.}

\begin{deluxetable*}{lcccccc}
\tablewidth{0pc}
\tablecaption{Simulation Summary\label{t:sims}}
\tablehead{
\colhead{Simulation}&
\colhead{Domain\tablenotemark{a}}&
\colhead{Resolution}&
\colhead{Orbits}&
\colhead{$\langle -B_x B_y\rangle/P_0$\tablenotemark{b}}&
\colhead{$\langle\rho v_x \delta v_y\rangle/P_0$\tablenotemark{b}}&
\colhead{$\langle B_y \rangle/\sqrt{P_0}$\tablenotemark{c}}
}
\startdata
S32R1Z4 &
$H \times 4H \times 4H$ &
$32/H$ &
300 &
0.012 &
0.0029 &
0.066\\
S64R1Z4 &
$H \times 4H \times 4H$ &
$64/H$ &
300 &
0.0075 &
0.0018 &
0.029\\
S128R1Z4 &
$H \times 4H \times 4H$ &
$128/H$ &
300 &
0.0076 &
0.0016 &
0.034\\
S32R1Z6 &
$H \times 4H \times 6H$ &
$32/H$ &
165 &
0.010 &
0.0024 &
0.074\\
S32R4Z4 &
$4H \times 4H \times 4H$ &
$32/H$ &
250 &
0.0082 &
0.0022 &
0.035\\
S64R4Z4 &
$4H \times 4H \times 4H$ &
$64/H$ &
160 &
0.0076 &
0.0019 &
0.040\\

\enddata
\tablenotetext{a}{$L_x \times L_y \times L_z$}
\tablenotetext{b}{Brackets denote temporal averages taken from
50 orbits onward and volume averages over whole domain.}
\tablenotetext{c}{Brackets denote temporal averages taken from
50 orbits onward and volume averages over innermost two scale
heights.}

\end{deluxetable*}

It is also instructive to look at the Fourier
transform of the induction equation.  Taking Fourier transforms
of the $x$ and $y$ components of (\ref{eq:induction}) we find
\be
\frac{1}{2}\frac{\partial|\bfourx|^2}{\partial t} = A_x + E_{z,y} +
E_{y,z},
\label{eq:fourx}
\ee
and 
\be
\frac{1}{2}\frac{\partial|\bfoury|^2}{\partial t} = S + A_y + E_{z,x} +
E_{x,z}.
\label{eq:foury}
\ee
We will focus on these two components as they appear to be the most
important for understanding the disk dynamo.

The definitions of the  terms on the right-hand sides of
(\ref{eq:fourx}) and (\ref{eq:foury}) are 
\be
S({\bm k}) = Re \left[\tilde{B}^{\ast}_{y}({\bm k}) \cdot 
\int B_x \frac{\partial V_{\rm sh}}{\partial x}
\exp{(-i {\bm k} \cdot {\bm x})} d^3 {\bm x}  \right],
\ee

\be
A_i({\bm k}) = -Re \left[\tilde{B}^{\ast}_{i}({\bm k}) \cdot \int V_{\rm sh} 
\frac{\partial B_i}{\partial y} \exp{(-i {\bm k} \cdot {\bm x})} d^3 {\bm x}
\right],
\ee

\be
E_{z,y}({\bm k}) = Re \left[\bconjx \cdot \int
\frac{\partial{\emf_z}}{\partial y}\exp{(-i {\bm k} \cdot {\bm x})} 
d^3 {\bm x}
\right],
\ee

\be
E_{y,z}({\bm k}) = - Re \left[\bconjx \cdot \int
\frac{\partial{\emf_y}}{\partial z}\exp{(-i {\bm k} \cdot {\bm x})} 
d^3 {\bm x}
\right],
\ee

\be
E_{z,x}({\bm k}) = - Re \left[\bconjy \cdot \int 
\frac{\partial{\emf_z}}{\partial x}\exp{(-i {\bm k} \cdot {\bm x})} 
d^3 {\bm x}
\right],
\ee
and
\be
E_{x,z}({\bm k}) = Re \left[\bconjy \cdot \int 
\frac{\partial{\emf_x}}{\partial y} \exp{(-i {\bm k} \cdot {\bm x})}
d^3 {\bm x}
\right],
\ee
where subscript $i$ in $A_i$ refers to either the $x$ or $y$
coordinate. The EMFs are defined as ${\bm \emf} = {\bm v_t} \times {\bf B}$,
with ${\bm v_t} = {\bm v} - {\bm v_{\rm sh}}$ and
\be
{\bm v}_{\rm sh} = \frac{\hat{\bm j}}{L_y L_z} 
\int \int v_y {\rm d}y {\rm d}z,
\ee
where $L_{y}$ and $L_{z}$ the size of the computational domain in the
$y-$ and $z-$directions.
The $A_x$ and $A_y$ terms are included for completeness,
but they are generally much smaller than the other terms so we will
not discuss them further.

These relations are similar to the transfer functions used in
\citet{fp07} and \citet{shb09}.  In fact, our definition of $S$ is
identical and if we sum $A_i$ over all three spatial dimensions, it
would equivalent to their definition of $A$. 
These authors expand
\bd
{\bm \nabla} \times ({\bm v_t}\times{\bm B})=
({\bm B \cdot \nabla}){\bm v_t}
-({\bm v_t \cdot \nabla}){\bm B}-({\nabla \cdot \bm v_t}){\bm B},
\ed
and perform similar Fourier analysis on the three right-hand side terms
individually, labeling them $T_{bv}$, $T_{bb}$, and $T_{\rm div}$,
respectively.
One drawback of this expansion is that terms such as $B_x \partial
v_x/\partial x$ are present, even though they do not contribute to the
evolution of ${\bm B}$, because they appear with opposites signs in both
$T_{\rm div}$ and $T_{bv}$.  Such terms can be quite large,
complicating the interpretation of $T_{bv}$, $T_{bb}$, and $T_{\rm
  div}$.  We prefer to leave the right hand sides in
terms of the EMFs.

For plotting purposes, we find it useful to normalize the quantities
on the right hand side of (\ref{eq:fourx}) and
(\ref{eq:foury}) with the power spectrum.  To differentiate them from
the unnormalized quantities, we will use lower case letters.  For
example, $e_{y,z}(k) \equiv 2 E_{y,z}(k)/(|\tilde{B}_x(k)|^2 \Omega)$.
This then constitutes the Fourier amplitude of the normalized rate of
field production of $B_x$ due to the vertical variation of $\emf_y$
The factor $\Omega$ has been introduced to make the quantities
dimensionless rates.

\begin{figure}[h]
\begin{center}
\hbox{
\psfig{figure=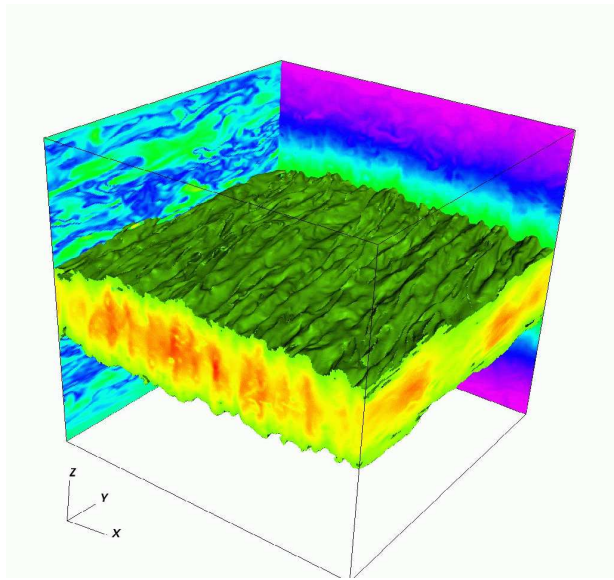,width=0.45\textwidth,angle=0}

}
\end{center}

\caption{
Isosurface (at $\rho=0.75$) and slices of the density at 250 orbits in
a domain of size $4H \times 4H \times 4H$ (S64Z4R4).  On the left face
of the domain a slice of the magnitude of the magnetic field is shown.
\label{f:3d}}
\end{figure}

\section{Results}
\label{res}

\subsection{Resolution Study}
\label{std}

Our primary goal is to test the robustness of sustained turbulence and
angular momentum transport in stratified shearing boxes with zero net
flux, and in addition, to characterize the properties of the
turbulence in this case.  Figure 1 is an image showing the three-dimensional
structure of the density at late time (250 orbits)
in a typical simulation, computed with a resolution of 64 grid zones
per scale height in an $4H \times 4H \times 4H$
domain.  Spiral density waves characteristic of all simulations of the
MRI in shearing boxes are evident in the density isosurfaces.

\begin{figure}[h]
\begin{center}
\hbox{
\psfig{figure=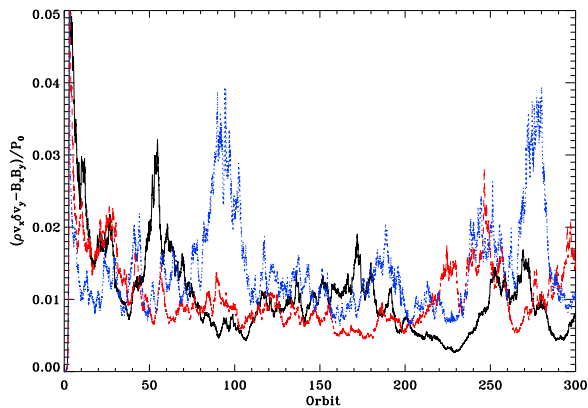,width=0.45\textwidth,angle=0}
}
\end{center}

\caption{
Sum of integrated Reynolds and Maxwell stresses as a function of time
in $H\times 4H\times 4H$ stratified shearing boxes. The curves
represent the $128/H$ (black, solid), $64/H$ (red, dashed), and $32/H$
(blue, dotted) resolutions.
\label{f:stress}}
\end{figure}

To investigate the convergence of the stress with numerical
resolution, we have performed a resolution study at 32, 64
and 128 grid zones per scale height in an $H \times 4H \times 4H$
domain (hereafter S32R1Z4, S64R1Z4, and S128R1Z4, respectively).
Since interest in MRI turbulence is driven primarily by its role in
angular momentum transport, we first focus on stress as a diagnostic.
Table \ref{t:sims} and Figure \ref{f:stress} summarize the behavior of
the stress as we vary the resolution.

The MRI turbulence contributes to the stress through the Maxwell
stress $-B_x B_y$ and the Reynolds stress $\rho v_x \delta v_y$, where
$\delta v_y$ is the $y-$component of the velocity with background shear
removed.  The domain and time average of these quantities are listed
in Table \ref{t:sims}.  We have normalized them by the initial
midplane pressure $P_0$.  Since the initial condition is in
hydrostatic equilibrium and magnetic pressure is always significantly
lower than gas pressure at the midplane (see e.g. Figure
\ref{f:st128}), this value of the midplane pressure does not evolve
significantly.  With this normalization they are roughly equivalent to
the $\alpha$ parameter of \citet{ss73}. The time average is carried
out from 50-300 orbits to exclude the transient period of enhanced
turbulence during and immediately after the linear growth phase of the
MRI.

Both contributions to the stress decrease as we increase resolution,
but the changes are much smaller when going from $64/H$ to $128/H$
than from $32/H$ to $64/H$, indicating convergence.  This should be
compared with the behavior observed in unstratified boxes
\citep[e.g.][]{san04,fp07,pcp07} in which total stress decreases by
factors of $\sim 2$ as resolution is increased by a factor of $2$.
The normalized total stress in the S128R1Z4 run is $0.0095$,
comparable to previous results for stratified domains with zero net
flux \citep{bra95,sto96,jyk09,si09}.  The Maxwell stress is 4-5 times
greater than Reynolds stress, which is slightly higher than, but
roughly consistent with previous results for stratified domains
\citep[e.g.][]{sto96}.  Similar values are also observed in
unstratified runs, where the result appears to be independent of field
geometry or flux, and depend mainly on the rate of shear
\citep{pcp06}.

Table \ref{t:sims} also includes the rms toroidal field, volume
averaged over the central two scale heights and time averaged from 50
orbits onward.  The rms field strengths are relatively weak, with
$\langle B_y\rangle^2 \sim 0.01 \langle B_y^2\rangle$, but may still
be dynamically important since the presence mean toroidal field in
unstratified simulations has been shown to enhance turbulence stresses
and energy densities \citep{hgb95}. In fact, the rms toroidal
field strength correlates well with the stress, although this may be
the by-product of stronger turbulence rather than a cause.

\begin{figure}[h]
\begin{center}
\hbox{
\psfig{figure=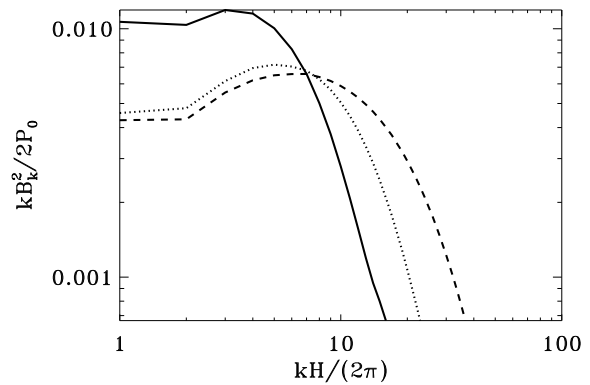,width=0.45\textwidth,angle=0}
}
\end{center}
\caption{
Comparison of magnetic energy density power spectra for $32/H$ (solid), $64/H$
(dotted), and $128/H$ (dashed) resolutions in $H\times 4H\times
4H$ stratified shearing boxes.  Power spectra have been averaged over spherical
shells of constant $k \equiv |{\bm k}|$ The $k^3$ normalization then makes
the y ordinate proportional to the 
fractional contribution to the total power per logarithmic interval in $k$.
The power spectra are time averaged from 50-300 orbits.
\label{f:pscomp}}
\end{figure}

Figure \ref{f:stress} shows the temporal variation of the total
stress.  There is considerable variability, with relatively long-lived
($\gtrsim 50$ orbit) departures from the mean.  In the S32R1Z4 run,
the $\sim 50$ orbit periods of enhanced stress contribute almost as
much to the average as the longer periods of `quiescent' stress. There are
similar long-term fluctuations in the higher resolution runs, but
these are generally smaller in amplitude and less important for
determining the overall mean.  Nevertheless, it is clear that one must
average over relatively long baselines to obtain a representative
value, making higher resolutions prohibitively expensive.

A power spectral analysis of the magnetic field also indicates
convergence, at least for the large scales where most of the power
resides.  Figure \ref{f:pscomp} shows the averaged PSD for the
S32R1Z4, S64R1Z4, and S128R1Z4 simulations.  We average over spherical
shells in $k$-space (see \S\ref{fft}) and in time, excluding the
first 50 orbits to avoid the initial transients. There is a
significant drop in power in going from the $32/H$ to $64/H$
resolution runs, combined with a shift in the peak of PSD to larger
$k$.  However, when going from $64/H$ to $128/H$, there is
significantly smaller drop in amplitude at most scales and a much
smaller shift in the peak wave number, also indicating convergence.

\begin{figure}[h]
\begin{center}
\hbox{
\psfig{figure=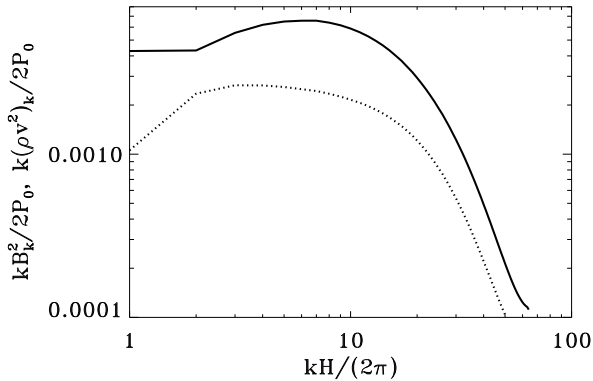,width=0.45\textwidth,angle=0}
}
\end{center}
\caption{
Comparison of magnetic (solid) and kinetic (doted) energy density power 
spectra in the $128/H$ resolution, $H\times 4H\times
4H$ stratified shearing box. 
\label{f:psmeke}}
\end{figure}

Although we focus on the magnetic energy density, a similar
convergence is observed in the PSD of the kinetic energy density.  We
show a comparison of the kinetic and magnetic energy density PSDs for
run S128R1Z4 in Figure \ref{f:psmeke}.  It is notable that power in the
magnetic field fluctuations exceeds that of the kinetic energy at all
scales.  This is in contrast to unstratified runs which typical show
greater power in the kinetic energy at the lowest $k$, with magnetic
energy dominating the power at higher $k$.  It also differs from
simulations of helically driven turbulence where the kinetic and
magnetic energy have comparable amplitude at all but the lowest
$k$ where magnetic energy dominates \citep{bra01}.

\begin{figure}[h]
\begin{center}
\hbox{
\psfig{figure=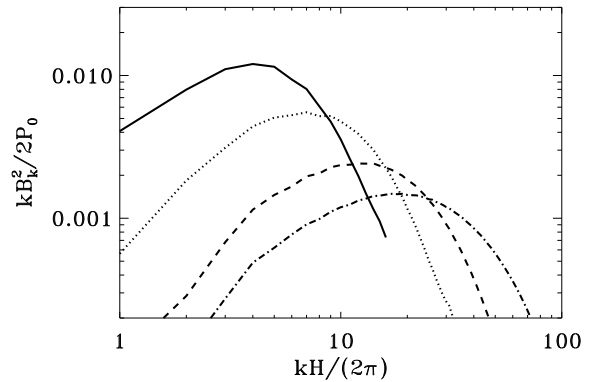,width=0.45\textwidth,angle=0}
}
\end{center}
\caption{
Comparison of magnetic energy density power spectra for $32/H$ (solid), $64/H$
(dotted), $128/H$ (dashed), and $256/H$ (dot-dashed) resolutions in 
$H\times 4H\times H$ unstratified shearing boxes.  Power spectra are time 
averaged from 60-100 orbits.  The spectra peak at $kH/(2\pi) =$ 4, 7, 14, and
17, for the $32/H$, $64/H$, $128/H$, and $256/H$ runs respectively.  This
roughly consistent with a  scaling $k_{\rm max} \simeq 2\pi (H \Delta)^{-1/2}$, where $\Delta$
is the spacing between grid zones.
\label{f:psun}}
\end{figure}

\begin{figure}[h]
\begin{center}
\hbox{
\psfig{figure=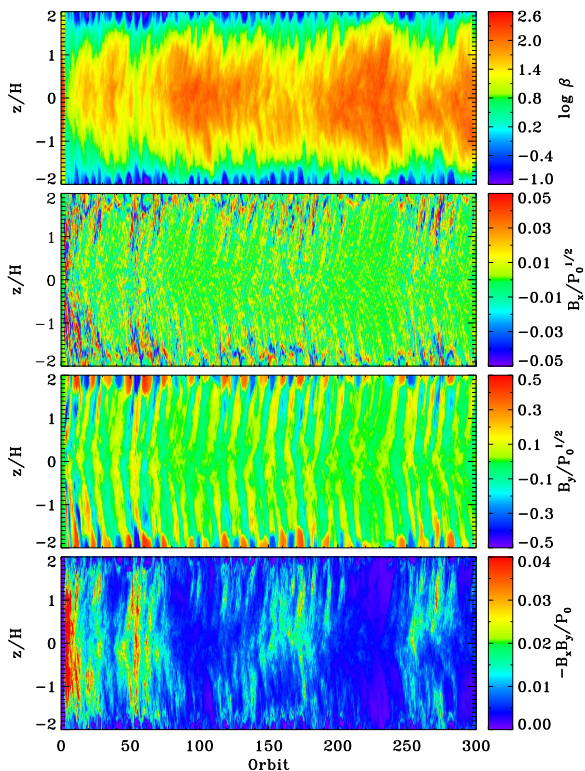,width=0.45\textwidth,angle=0}
}
\end{center}
\caption{
Spacetime plot for the $128/H$ resolution run in a $H\times 4H\times
4H$ stratified shearing box.  From top-to-bottom the panels show the
horizontally averages of plasma $\beta$, the normalized radial and
toroidal components of the magnetic field (respectively), and the
Maxwell stress as a function of height above the midplane.
\label{f:st128}}
\end{figure}

The convergent behavior of the stratified runs should be contrasted
with that of the unstratified simulations shown in Figure
\ref{f:psun}.  This plot shows the PSD for four unstratified runs with
resolutions of $32/H$, $64/H$, $128/H$, and $256/H$ in $H \times 4H
\times H$ shearing boxes.  Each factor of two increase in resolution
results in a decrease by nearly a factor of two in the integrated
power. There is also a shift in the peak wavenumber to larger $k$ as
resolution increases. This lack of convergence is almost identical to
the that found by previous authors \citep{fp07,gua09,shb09}.  It seems
that both the power and characteristic scale of the turbulence are set
by the domain resolution.  Adding stratification appears to
fundamentally change the dynamics and provides a characteristic scale
and amplitude of the turbulence which is independent of the
resolution.  This could be related to the different mechanisms that
lead to the saturation of the MRI in the presence of stratification,
perhaps associated with the development, and subsequent buoyant rise,
of large scale magnetic fields \citep{pcp07}.

In Figure \ref{f:st128} we show spacetime diagrams for several
variables associated with the magnetic field.  For the sake of
brevity, we focus on magnetic quantities, which appear to play the
dominant role, as suggested by the PSD analysis above.  The spacetime
plots are generated by averaging over the $x$ and $y$ coordinates at each
height in the domain every quarter of an orbit.  The top panel of
Figure \ref{f:st128} shows $\beta = 2 \langle P\rangle/\langle
B^2\rangle$, where the angle brackets denote horizontal averages.  As
noted previously, magnetic pressure remains relatively weak near the
midplane, but dominates in the surface regions ($|z| \gtrsim 1.5 H$).

The middle panels show the normalized $x$ and $y$ components of the
magnetic field.  The $B_y$ component is considerably larger than $B_x$
and they are negatively correlated.  The symmetry of the boundary
conditions and the initial conditions require the vertical average of
these quantities to be zero.  Nevertheless, there are localized
regions of net field that, under the effects of buoyancy, trace out
curved trajectories in the spacetime plot.  This is similar to other
`butterfly diagrams' seen in previous shearing box calculations of
stratified domains \citep{bra95,sto96,tur04,jyk09,si09}.  Consistent
with previous simulations, the polarity is usually even about the
midplane and, at fixed height, oscillates quasiperiodically on time
scales of $\lesssim 10$ orbits.

Within the inner three scale heights, the horizontally averaged net
field is a sum over a fluctuating $\bm B$ field so that $\langle B_y
\rangle^2$ is much less than $\langle B_y^2 \rangle$ and similarly for
$B_x$.  As these regions of net field buoyantly rise, the ratio of
$\langle B_y \rangle^2/\langle B_y^2 \rangle$ increases.  Near the
vertical boundaries magnetic dominated regions of rather uniform ${\bm
  B} \sim B_y {\hat{\bm j}}$ develop.  Their presence is very likely
related to our assumption of periodicity in the vertical boundary
condition, so they are likely not physically relevant to real
accretion flows.  The degree to which these regions affect the
dynamics is discussed further in \S\ref{tall}.

The bottom panel of Figure \ref{f:st128} shows the Maxwell stress.  In
addition to the temporal fluctuations, there is also considerable
variation with height. The middle three scale heights dominate and
regions of greatest stress tend to be found off the midplane.  The
stress is weakest in the magnetically dominated regions very near the
boundary, and is even slightly negative in some places (although the
colorbar only goes to zero). A striking result is the correlation of
regions of stronger than average $\langle B_y \rangle$ with regions of
larger than average stress.  This lends support to the idea that the
presence of a mean toroidal field leads to enhanced turbulent stresses
and energy densities.  Although not shown, we note that the spacetime
plot of the Reynolds stress is very similar to the Maxwell stress and
the two are well correlated in both $z$ and time.

\begin{figure}[h]
\begin{center}
\hbox{
\psfig{figure=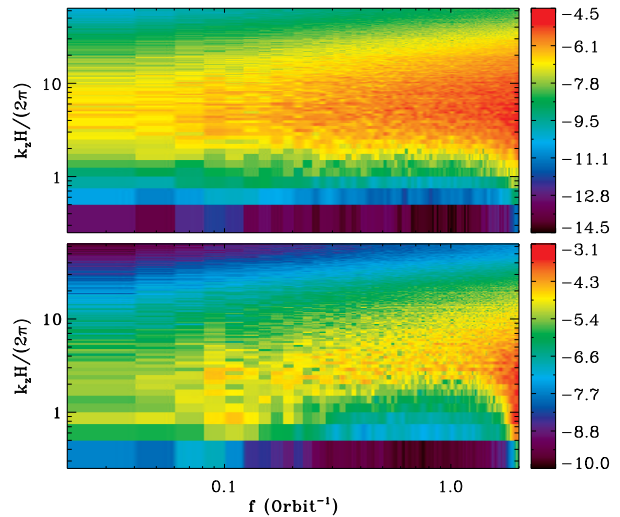,width=0.45\textwidth,angle=0}
}
\end{center}
\caption{
Magnetic energy power spectra as function of both $k_z$ and $f$,
where $f$ is the frequency in the time domain.  The top and bottom
panels correspond to the amplitudes of the $B_x$ and $B_y$ contributions
the magnetic energy density, respectively.  A detailed description
is provided in the text (\S\ref{std}).
\label{f:kvsf}}
\end{figure}

Figure \ref{f:st128} shows there is clearly a significant amount of
large scale and long timescale structure in space and time
coordinates, respectively.  To better understand and quantify this
structure, we perform a complimentary Fourier analysis in both space
and time.  Since we are interested in the structure of the
horizontally averaged box, we focus on vertical $k$-vectors with ${\bm
  k}=k_z \hat{\bm z}$.  Every one-quarter of an orbit, we compute the
discrete Fourier transform $B^2(k_{z}) \equiv 4\pi k_{z}^2 |\tilde{B}
(k_{z})|^2$, which is analogous to the $B^2_k$ defined in \S\ref{fft},
but with $k_z$ replacing $k$.  Note that the two quantities can differ
significantly since the Fourier amplitudes are far from isotropic in
$k$-space.  For each $k_z$, we Fourier transform in time to obtain
$B^2(k_{z},f)$ where $f$ is the time domain frequency.  We divide the
data into five 50 orbit time series (between 50 and 300 orbits),
Fourier transform each separately, and then average.  Although we lose
access to the longest timescales, the averaging reduces the variance
in the resulting power spectra, which are plotted in Figure
\ref{f:kvsf}.

We plot both $\log[f k_z B^2_x(k_{z},f)/(2 P_0)]$ (top panel) and
$\log[f k_z B^2_y(k_{z},f)/(2 P_0)]$ (bottom panel).  Note that the
horizontal average $\langle B_z \rangle$ is conserved (at zero) to
round-off error, so there is no physical information in the $z$
component for vertical wavevectors. We have multiplied by both $k_z$
and $f$ before taking the logarithm.  Since we use a logarithmic scale
for both the $k_z$ and $f$ axes, this weights each pixel so that its
color scales linearly with it contribution to the overall power (i.e.
in the same sense that $\nu F_\nu$ is commonly used in astrophysics).

There are significant differences in the morphologies of the $B_x$ and
$B_y$ power spectra.  For large spatial scales (small $k_z$), both
$B_x$ and $B_y$ show a double peaked profile with significant power at
large ($\sim 10$ orbit) and small ($\lesssim 1$ orbit) timescales,
although the small scales are subject to aliasing.  The dip at
intermediate scales is somewhat more pronounced and persists to
somewhat larger $k_z$ for $B_y$ than for $B_x$.  As we shift to larger
$k_z$, the peak in $B_x$ broadens significantly and the dip goes away
entirely.  For $B_y$ there is a locus of maximum power which shifts to
higher $k_z$ as $f$ increases from about 0.1 to 1 cycles per orbit.

\subsection{Vertically Extended  Domains}
\label{tall}

Due to the low densities and high magnetic field strengths near the
vertical boundary that arise from stratification, increasing the
vertical extent of the domain is particularly computationally
expensive.  Therefore, we performed our resolutions study in boxes
with four vertical scale heights, which seemed suitable to get a
separation between the midplane dynamics and the vertical boundary.
In order to confirm that our results are not strongly dependent on
this choice, we have repeated our $32/H$ resolution run with six
vertical scale heights (hereafter S32R1Z6).  As one can see in Table
\ref{t:sims}, the resulting volume average of the stress in the two
$32/H$ simulations is in reasonable agreement, although slightly
smaller in the S32R1Z6 run.

\begin{figure}[h]
\begin{center}
\hbox{
\psfig{figure=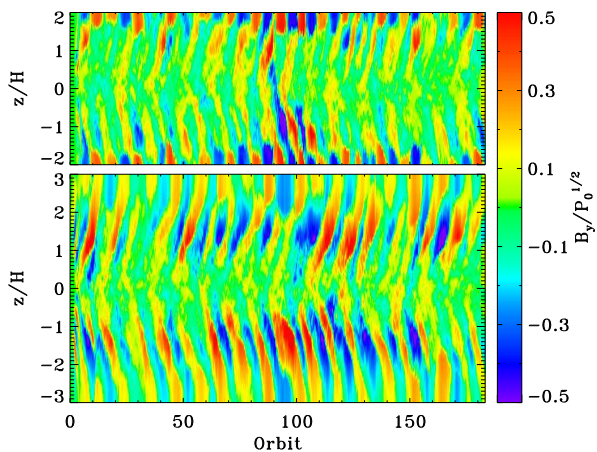,width=0.45\textwidth,angle=0}
}
\end{center}
\caption{
Spacetime plot comparing the toroidal component of the magnetic field
in stratified shearing boxes with $32/H$ resolution.  The top and
bottom panels show simulations with four and six scale height vertical
extent (respectively).
\label{f:st_by}}
\end{figure}

\begin{figure}[h]
\begin{center}
\hbox{
\psfig{figure=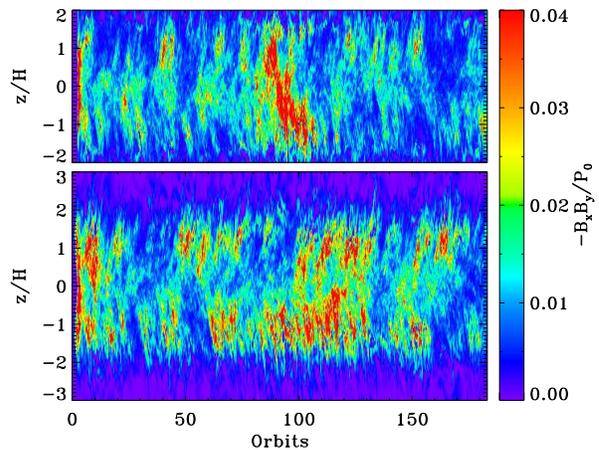,width=0.45\textwidth,angle=0}
}
\end{center}
\caption{
Spacetime plot comparing the Maxwell stress in stratified shearing
boxes with $32/H$ resolution.  The top and bottom panels show
simulations with four and six scale height vertical extent
(respectively).
\label{f:st_max}}
\end{figure}

One downside to choosing vertically periodic boundary conditions, is
the buildup of strongly magnetized regions with uniform $B_y$ near the
boundaries.  Although these regions don't contribute significantly to
the angular momentum transport, they are likely unphysical so one
might worry that they feedback on the dynamics closer to the midplane.
In order to get a better sense of their effect on the flow, we have
plotted spacetime diagrams comparing the S32R1Z4 and S32R1Z6
simulations in Figures \ref{f:st_by} and \ref{f:st_max}.

Figure \ref{f:st_by} shows the $y$ component of the magnetic field.
In a larger domain, there are still regions of rather uniform $B_y$
near the vertical boundaries.  They are somewhat more extended in
height but with a slightly weaker net field.  The regions of net
$B_y$ generated near the midplane can buoyantly rise to larger heights
before interacting with the boundary region.  Since the horizontally
averaged field tends to increase as the fluid rises, this lead to
further enhancement of the field strength over those found in the
smaller domain.

Figure \ref{f:st_max} shows the Maxwell stress in the two runs.  In
the central four scale heights the two plots look very similar,
suggesting that the four scale height runs are yielding a fairly
robust estimate for the angular momentum transport.  Regions of
enhanced Maxwell stress are again correlated with regions of strong
net $B_y$.  Similarly, the Maxwell stress is generally larger
in the inner four scale heights than in the S32R1Z4 run.  Note that
the volume weighted average stresses in Table \ref{t:sims} are lower
for S32R1Z6 than for S32R1Z4 because we are averaging over the whole
box, including the regions of weak stress near the boundaries.  If we
restrict the averaging to the inner two scale heights for both the
S32R1Z6 and S32R1Z4 runs, the time averaged Maxwell stress in the
S32R1Z6 simulation is greater by about 10\%.

\subsection{Radially Extended Domains}
\label{wide}

For the sake of computational expediency, we have performed our
resolution study on domains with only one scale height in the radial
direction.  This has traditionally been the box size employed in most
shearing box computations, mostly due to CFL constraints on the
timestep which are imposed by the background shear.  Using Athena's
orbital advection scheme (discussed in \S\ref{meth}), we can consider
larger domains to examine the effect of this choice on our results. We
have computed shearing boxes with $4 H \times 4H \times 4H$ domains at
$32/H$ and $64/H$ resolution (hereafter S32R4Z4 and S64R4Z4,
respectively).

\begin{figure}[h]
\begin{center}
\hbox{
\psfig{figure=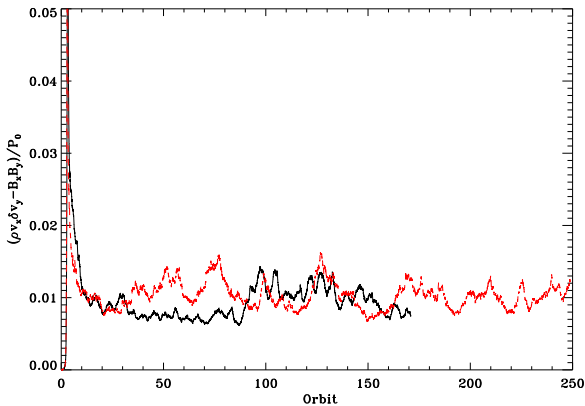,width=0.45\textwidth,angle=0}
}
\end{center}
\caption{
Sum of box integrated Reynolds and Maxwell stresses as a function of
time in $4H\times 4H\times 4H$ stratified shearing boxes. The curves
represent the $64/H$ (black, solid) and $32/H$ (red, dashed)
resolutions.
\label{f:stress4}}
\end{figure}

We plot the evolution of the total stress in these two simulations in
Figure \ref{f:stress4} and the normalized, time and volume averaged
stresses are listed in Table \ref{t:sims}.  The mean values of the
stress are very similar to each other and also to those found at
higher resolutions runs in the smaller boxes (S64R1Z4 and S128R1Z4).
This suggest that convergence is occurring at even lower resolution
than in the smaller domain computations.  The time evolution
differs from that seen in Figure \ref{f:stress} in that the amplitude
of fluctuations is much lower. 

\begin{figure}[h]
\begin{center}
\hbox{
\psfig{figure=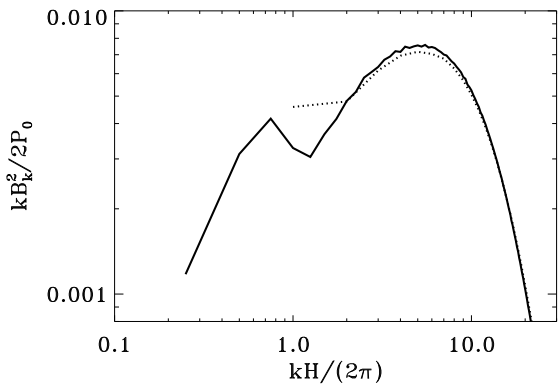,width=0.45\textwidth,angle=0}
}
\end{center}
\caption{
Comparison of magnetic energy density power spectra in 
stratified shearing boxes with $64/H$
resolution for domains with $L_x=4H$ (solid) and $L_x=H$ (dotted).
The larger radial extent in the former allows one to isotropically average
shell to lower $k$.
\label{f:psasp}}
\end{figure}

In Figure \ref{f:psasp} we compare the PSDs from simulations with
different radial extent but the same resolution (S64R1Z4 and S64R4Z4).
The PSDs are very similar at all but the lowest $k$.  At low $k$
the differences are accounted for in part by our shell averaging
scheme.  Since the larger box is a $4H$ cube, we can isotropically
average shells all the way to $k=\pi/(2H)$.  Since we can not do this
average isotropically with smaller box, some of this low $k < 2\pi/H$
power is included in the $k = 2\pi/H$ bin.  Overall, the PSDs seem to
be rather independent of the aspect ratio, consistent with nearly
identical values for the the box integrated stresses.

\begin{figure}[h]
\begin{center}
\hbox{
\psfig{figure=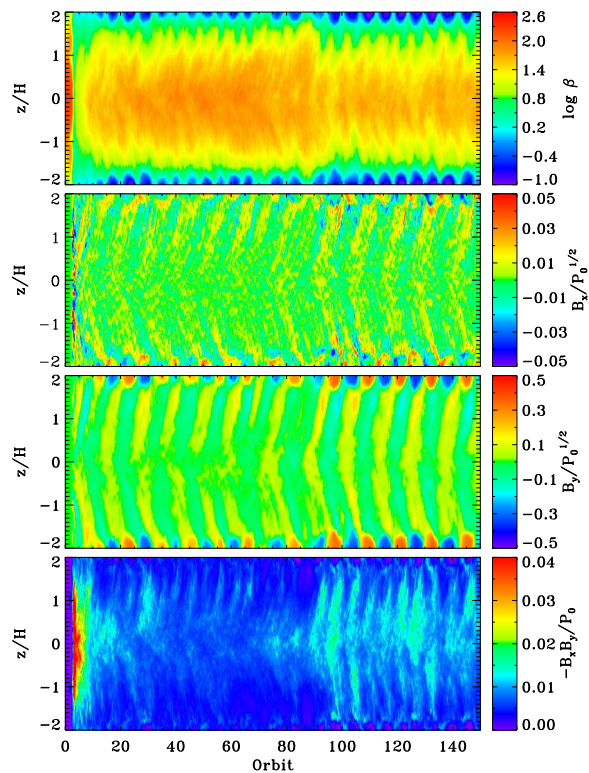,width=0.45\textwidth,angle=0}
}
\end{center}
\caption{
Spacetime plot for the $64/H$ resolution run in a $4H\times 4H\times
4H$ stratified shearing box.  From top-to-bottom the panels show the
horizontally averages of plasma $\beta$, the normalized radial and
toroidal components of the magnetic field (respectively), and the
Maxwell stress as a function of height above the midplane.
\label{f:st64r4}}
\end{figure}

Figure \ref{f:st64r4} shows the spacetime diagram for the S64R4Z4 run.
Overall, it is rather similar to S128R1Z4 spacetime diagram in Figure
\ref{f:st128}.  There are long timescale ($\gtrsim 50$ orbit)
fluctuations in the Maxwell stress, as well as $\sim 10$ orbit
quasi-periodic variations in $B_y$ and Maxwell stress which
are qualitatively similar to those in Figure \ref{f:st128}.  However,
the amplitude of fluctuations is generally smaller in the larger box,
consistent with the lower amplitude fluctuations in the total stress
found in Figure \ref{f:stress4}.

In order to better understand the lower fluctuation amplitudes, we
have split the larger domain into four sub-domains of one scale height
each in the radial direction.  We find that the volume averaged
stresses in each subdomain are highly correlated with each other.
Therefore, the lower amplitude is not simply the result of
``averaging'' over several independent boxes, but appears to be a
global property of a correlated domain.  This behavior is somewhat
surprising in light of the results of \citet{gua09}, which show that
the turbulence decorrelates on scales $\gtrsim H$ in unstratified
boxes.

To investigate the correlation in stratified boxes, we have followed 
\citet{gua09} and calculated the trace of the two-point correlation 
of the magnetic
field
\be
\xi_B=\langle\delta B_i({\bm x}) \delta B_i({\bm x}+\Delta {\bm x})\rangle
\label{eq:corr}
\ee
where $\delta B_i=B_i -\langle B_i \rangle$ and there is an implied
summation over $i$.  We computed the average in two ways: using the
full domain and using only the innermost two scale heights.  The two
different procedures give different results for the correlations in the
$x-y$ plane at large separations, since full domain average is
dominated by the magnetically dominated regions near the vertical
boundary.  Since these are likely unphysical, we report only the two scale
height average.

\begin{figure}[h]
\begin{center}
\hbox{
\psfig{figure=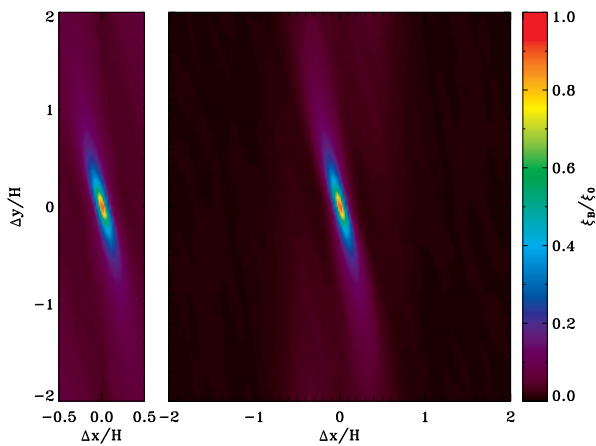,width=0.45\textwidth,angle=0}
}
\end{center}

\caption{
Two-point correlation functions for the magnetic field in the $x-y$
plane.  The axis labels $x$ and $y$ refer to the $\Delta x$ and
$\Delta y$ implicit in (\ref{eq:corr}).  We have normalized the
$\xi_B$ by it's maximum value which occurs at $\Delta x=\Delta y=0$.
The left and right panels correspond to the S64R1Z4 and S64R4Z4
calculations, respectively.
\label{f:corr}}
\end{figure}

We compute the correlation for 21 and 18 evenly spaced snapshots for
for the S64R1Z4 (50-300 orbits) and S64R4Z4 (50-150 orbits)
simulations, respectively.  Although correlation at small separations
is relatively consistent from snapshot to snapshot, there is
significant variation at larger scales.  Therefore, we average over
all snapshots to get a mean correlation for each run.  We plot the
results of the two scale height average for the $x-y$ plane in Figure
\ref{f:corr}.  For both simulations we normalize $\xi_B$ by their
maximum values, $\xi_0$, which agree to within 10\%.  Our results are
qualitatively consistent with those of \citet{gua09} in that we find
comparable tilt angles $\theta_t$, which is the angle between the
major axis of the correlation and the azimuthal axis.  The correlation
is very similar in both simulations, although the tilt angle from the
S64R4Z4 calculation is slightly smaller ($\theta_t \sim 15^\circ$
rather than $18^\circ$).

\begin{figure}[h]
\begin{center}
\hbox{
\psfig{figure=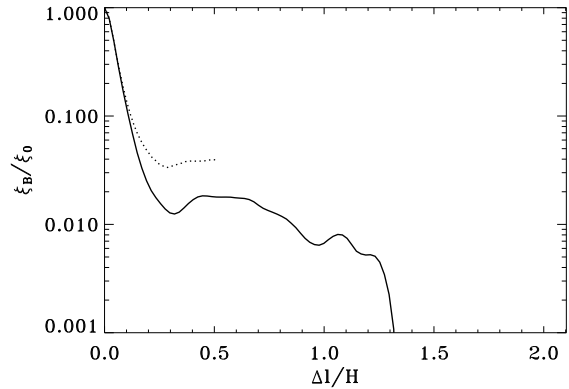,width=0.45\textwidth,angle=0}
}
\end{center}

\caption{
Magnetic field correlation along the minor axis in the $x-y$ plane,
plotted for S64R4Z4 (solid) and  S64R1Z4 (dotted).  For the horizontal
axis, $\Delta l$ is
the displacement from $\Delta x=\Delta y=0$ along the minor axis
of the correlation.
\label{f:axis}}
\end{figure}

We also plot the correlation along the minor axis (defined by
$\hat{\bm x} \cos \theta_t + \hat{\bm y} \sin \theta_t$) in Figure
\ref{f:axis}.  Again, the core of the correlation at small separations
is nearly identical in the two simulations and consistent with the
results of \citet{gua09}.  Near $\Delta l \sim 0.1 H$ the slope of the
curve from the S64R1Z4 run flattens and the large scale correlation
plateaus at a value of $\xi_B \simeq 0.04 \xi_0$.  The S64R4Z4 curve
declines further, also flattens with $\xi_B \simeq 0.01 \xi_0$ until
$\Delta l \sim 1.3 H$ where it drops nearly to zero.  

As equation (\ref{eq:corr}) requires, we have subtracted the mean
field which is generally non-zero when only the inner two scale
heights are considered (see Table \ref{t:sims}), but is identically
zero when using the whole domain.  These mean fields can be
substantial and would dominate the large scale correlation if
included.  It seems that these fields are sufficient to enforce
relative uniformity in the magnetic energy and Maxwell stress
throughout the four scale height domain. This uniformity may disappear
for sufficiently large domains, and we see some suggestion of this in
a calculations at $32/H$ resolution where we have compared $8H \times
8H \times 4H$ and $4H \times 4H \times 4H$ domains.  Although we find
that in both cases the Maxwell stress and magnetic energy density in
one scale height wide subdomains remain correlated, they show greater
variance in the the eight subdomains of the large box than in the four
subdomains of the smaller box.

\subsection{Domains with Finite Dissipation}
\label{dissip}

It has been shown that the behavior of MRI turbulence in unstratified
shearing boxes depends on the nature of the dissipation
\citep{sim98,fsh00,fro07,ll07,sh09}. Simulations with explicit
dissipation yield different results than those with only numerical
dissipation, and the strength and evolution of the turbulence depends
on both the viscosity $\nu$ and resistivity $\eta$, or equivalently
the the Reynolds number $Re \equiv c_s H/\nu$ and the magnetic
Reynodls number $Rm \equiv c_s H/\eta$.  Specifically, it has been
shown that $Re$ and $Rm$, or alternatively the magnetic Prandtl number
$Pm=Rm/Re$, determine whether turbulence is sustained in zero net
flux, unstratified simulations \citep{fro07,sh09}. 

To see if these results hold in stratified domains, we perform three
simulations with differing values of viscosity and resistivity: Re=800
with Pm=4 (hereafter Re800Pm4), Re=800 with Pm=2 (Re800Pm2), and
Re=1600 with Pm=2 (Re1600Pm2).  Examination of Figure 11 of
\citet{fro07} or Table 1 of \citet{sh09} show that none of these would
sustain turbulence in an unstratified domains with no net field,
regardless of whether Zeus or Athena is used for the simulations.  We
have confirmed these results for unstratified domains with our own
Athena calculations.

\begin{figure}[h]
\begin{center}
\hbox{
\psfig{figure=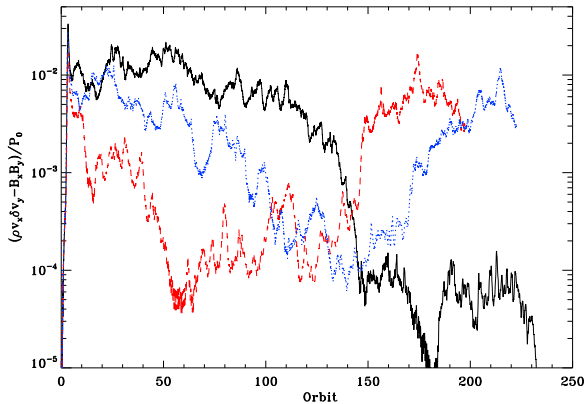,width=0.45\textwidth,angle=0}
}
\end{center}
\caption{
Sum of box integrated Reynolds and Maxwell stresses as a function of
time in stratified shearing boxes with explicit
dissipation. The curves represent computations with Re=800, Pm=4 
(black, solid); Re=1600, Pm=2  (blue, dotted)
Re=800, Pm=2 (red, dashed).
\label{f:stressdis}}
\end{figure}

Figure \ref{f:stressdis} shows the stress as a function of time for
the three simulations with a logarithmic vertical scale.  The behavior
of Re800Pm4 and Re1600Pm2 is rather different from the unstratified
domains where the simulations decay rapidly to zero on timescales of
10-20 orbits after the initial linear growth of the MRI.  Even
Re800Pm2, which drops rapidly to a dimensionless stress of $\sim
10^{-3}$ and then $\sim 10^{-4}$ in the stratified domain, decays much
more rapidly and continues to even lower values in the unstratified
domain.  The behavior of the stratified domains is also considerably
more complicated.  Turbulence never decays away completely, but
vigorous turbulence is not sustained in any of the calculations for
longer than 100 orbits.  The amplitude of variability is large, and
turbulence decays slowly on timescales of hundreds of orbits.  The
Re1600Pm2 and Re800Pm2 runs both show a recovery nearly to peak values
after spending over 100 orbital periods in stagnation or slow decay!

Using the criteria of \citet{fro07}, we would probably have labeled
the Re800Pm4 run as having sustained turbulence (over the first 100
orbits, which is the baseline used there), the Re1600Pm2 run as
marginal, and the Re800Pm4 as either marginal or not having sustained
turbulence, although the complex variability of the stratified runs
makes this somewhat subjective.  Figure 11 of \citet{fp07} maps out a
locus of sustained turbulence in Re -- Pm space, and it's notable that
Re800Pm4 and Re1600Pm2 simulations are on the cusp of showing
sustained turbulence while Re800Pm2 more firmly in the non-turbulent
regime.  Therefore, it would seem that stratification slightly
increases the parameter space for which sustained turbulence is
possible, but does not qualitatively alter the conclusion that
turbulence dies out for sufficiently low Pm or sufficiently high
Re.

\begin{figure}[h]
\begin{center}
\hbox{
\psfig{figure=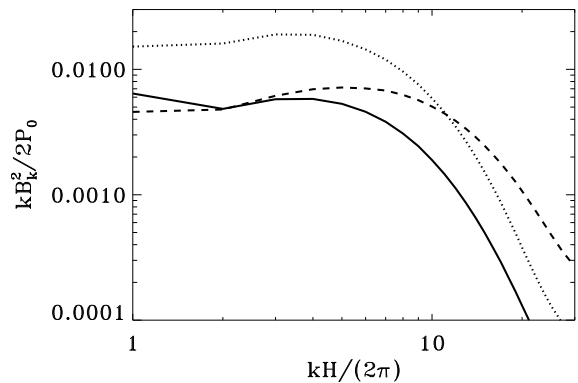,width=0.45\textwidth,angle=0}
}
\end{center}
\caption{
Comparison of magnetic energy density power spectra for computations 
with (solid, dotted) and without (dashed) explicit dissipation.  The explicit
dissipation curves correspond to computations with Re=1600, Pm=2 (solid) and
Re=800, Pm=4 (dotted). Power spectra are time averaged
from 25-100 orbits (with dissipation) or from 50-300 orbits (without dissipation).
\label{f:psdis}}
\end{figure}

It is suggestive that all three sets of dissipation terms show
sustained turbulence in unstratified boxes once a net toroidal field
is imposed \citep{sh09}.  As noted previously, it is conceivable that
main impact of stratification is the production of toroidal field,
which then lead to enhanced turbulence.  The turbulence in the
stratified runs is significantly less vigorous, but this may be
consistent with the rms field strengths in the stratified simulations
being much weaker than the toroidal fields considered by \citet{sh09},

In Figure \ref{f:psdis} we plot the time average power spectra of the
magnetic energy density from Re800Pm4 and Re1600Pm2, including S64R1Z4
for comparison.  Since the magnetic energy in the Re1600Pm2 drops
rapidly to a low amplitude, we have elected to exclude it. Of course,
the amplitude of the power spectrum depends on the interval used in
the time average, which is 25-100 orbits.  The Re800Pm4 and Re1600Pm2
power spectra are similar in shape, falling off somewhat more rapidly
than S64R1Z4 as $k$ increases.

The power in Re800Pm4 in exceeds that in Re1600Pm2 at all $k$.  Note
that these two calculations have the same resistivity but Re800Pm4 has
a higher viscosity, so it has higher amplitude despite having
larger overall dissipation (although this conclusion depends on the
interval used for the comparison).  Similar behavior is also observed
in two-dimensional simulations of MRI driven turbulence with a
vertical magnetic field and non-zero viscosity described in
\citet{ms08}.  In some of their runs saturation is not achieved since
the turbulent stresses increase with time until the end of the runs.
The fact that the magnetic field generated by the MRI can reach high
amplitudes could be due to the viscous quenching of Kelvin-Helmholtz
parasitic modes \citep{gx94,pg09}.  Although plausible, it is less
obvious that this process is responsible for the similar behavior
observed in the simulations with stratification and non-mean magnetic
flux presented here.

\section{Discussion}
\label{disc}

The numerical experiments presented here motivate two related
questions: Why do the stratified simulations converge when the
unstratified calculations clearly do not? What is the source of the
dynamo cycles observed in the large scale fields?  Answering the
former requires detailed comparison with unstratified simulations, and
will be the focus of future work.  For the moment, we focus primarily
on describing the large scale dynamo.

First, we compare with the discussion of \cite{bra95}, who noted the
similarities of their stratified results with an $\alpha-\Omega$ dynamo
model. They showed that the azimuthal EMF $\emf_y =(\delta {\bm v}
\times {\bm B})_y$ was correlated with the azimuthal field $B_y$ in a
manner that leads to growth in the radial $B_x$.  Coupled to the
shear, which regenerates $B_y$ from $B_x$, this describes a simple
dynamo.

Our results are in good agreement with their Equation (21).  For the
S128R1Z4 simulation, we find 
\be 
\langle \emf_y \rangle \simeq (-3, 5) \times 10^{-3} \langle 
B_y\rangle \langle {\delta v}^2 \rangle^{1/2}, 
\ee 
where the values in parentheses correspond to volume averages one
scale height above and below the midplane, respectively.  They also
report a correlation of $\langle \emf_x \rangle$ with $\langle B_y
\rangle$ in their Equation  (22).  Again, our results are qualitatively
similar to theirs with
\be 
\langle\emf_x\rangle \simeq
(-1, 1) \times 10^{-2} \langle B_y\rangle \langle{\delta
  v}^2\rangle^{1/2}, 
\ee 
above and below the midplane.  The first correlation confirms that
there is a mechanism for regenerating the poloidal field from a
toroidal field.  The second relation suggests that
$\langle\emf_x\rangle$ generally acts to reduce the magnitude of
$\langle B_y\rangle$ at the midplane, and is (at least partially) the
result of buoyancy, as \cite{bra95} discuss.

\begin{figure}[h]
\begin{center}
\hbox{
\psfig{figure=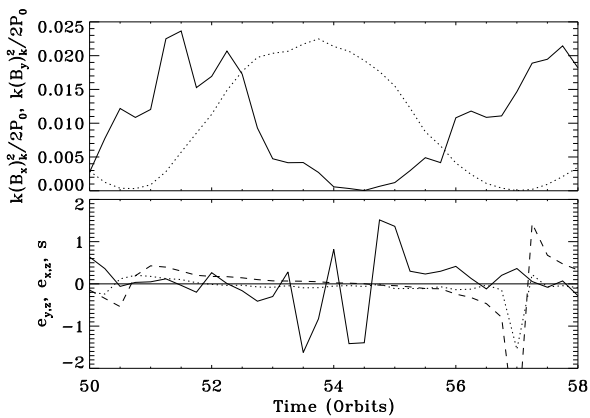,width=0.45\textwidth,angle=0}
}
\end{center}
\caption{
Power spectra (top) and normalized EMFs (bottom) for
a single oscillation period of S128R1Z4.  The top panel shows
PSDs $k|\tilde{B}_x(k_0)|^2/(2 P_0)$ (solid) and 
$k|\tilde{B}_y(k_0)|^2/(2 P_0)$ (dotted), where the former has
been multiplied by a factor of 400 for plotting convenience.
The bottom panel shows $e_{y,z}(k_0)$ (solid), $e_{x,z}(k_0)$ (dotted), and
$s(k_0)$ (dashed), which are defined in \S\ref{fft}.
\label{f:emft}}
\end{figure}

It's instructive to examine this further with the Fourier analysis
methods described in \S\ref{fft}.  We focus on the large scale field
and consider the smallest vertically oriented vector ${\bm k} = k_0
\hat{\bm z} =2 \pi/L_z \hat{\bm z}$.  Since, $k_x=k_y=0$, this term
represents the Fourier amplitude of ``horizontally averaged''
quantities on the largest vertical scale.  In Figure \ref{f:emft} we
plot time variation of magnetic fields and EMFs over a single dynamo
cycle for this choice of $\bm k$. The top panel shows the Fourier
amplitudes of magnetic energy densities $|\tilde{B}_x(k_0)|^2$ (solid)
and $|\tilde{B}_y(k_0)|^2$ (dotted) for this wave vector.  We have
multiplied $|\tilde{B}_x(k_0)|^2$ by a factor of 400 to plot both on
the same scale.  There is considerable variation from cycle to cycle,
and Figure \ref{f:kvsf} shows that these oscillations are only
quasiperiodic with a broad range of power for periods near 10 orbits,
depending somewhat on the wavenumber used for the analysis.
Nevertheless, this example is typical in that the curves are out of
phase with a more uniform variation in $|\tilde{B}_y(k_0)|^2$ than in
$|\tilde{B}_x(k_0)|^2$.

The bottom panel shows the corresponding right hand side quantities in
(\ref{eq:fourx}) and (\ref{eq:foury}), which, along with
numerical dissipation, drive the evolution of the Fourier amplitudes.
We normalize these quantities with the power spectra as described in
\S\ref{fft}, e.g.  $e_{y,z}(k_0) = 2 E_{y,z}(k_0)
/(|\tilde{B}_x(k_0)|^2 \Omega)$.  We plot $e_{y,z}$ (solid), $e_{y,z}$
(dotted), and $s$ (dashed).  Note that $e_{z,y}(k_0)$ and
$e_{z,x}(k_0)$ are zero for $k_0$ because of the periodic boundaries.
The shear term $s$ primarily drives the variation of
$|\tilde{B}_y(k_0)|^2$, flipping sign as $|\tilde{B}_x(k_0)|^2$ goes
to zero.  In contrast, $e_{x,z}$ is generally negative, acting as
turbulent resistivity.  The $e_{y,z}(k_0)$ term is more erratic,
frequently flipping sign over a single cycle, but the net effect is an
overall oscillation of $|\tilde{B}_x(k_0)|^2$ over $\sim 7$ orbital
periods.

In many respects, the behavior we see in the stratified simulations is
similar to that observed in the unstratified, zero-net flux
calculations of \citet{lo08}.  Using an incompressible spectral code,
they find dynamo cycles with a $\sim 5$ orbit periodicity.  This is
similar to the oscillations in our stratified runs where rms power on
large scales is broadly distributed on times scales $\sim 6-10$ orbits
(see Figure \ref{f:kvsf}).  The normalized quantities plotted in
Figure \ref{f:emft} are equivalent\footnote{Their notation reverses
  the definition of $x$ and $y$ from that used here: their $B_x$ and
  $B_y$ are the toroidal and radial field components, respectively.}
to their Equation (16).  Comparison of Figure \ref{f:emft} with Figs.
4 and 5 in their paper, show that the behavior of the EMFs during
oscillations are also quite similar, suggesting that a common (or, at
least, related) mechanism may be responsible for these oscillations.
This motivates a more detailed comparison between unstratified and
stratified runs in future work.

\begin{figure}[h]
\begin{center}
\hbox{
\psfig{figure=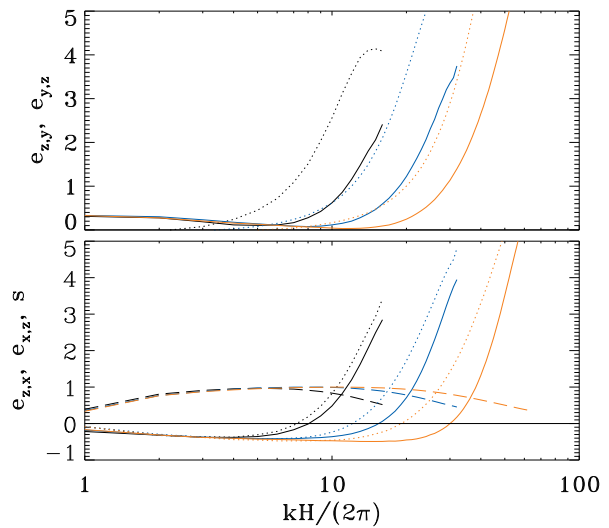,width=0.45\textwidth,angle=0}
}
\end{center}
\caption{
Time averaged and normalized EMFs for $x$ (top panel) and $y$ (bottom
panel) components of the induction equation.  The curves correspond to
the S32R1Z4 (black), S64R1Z4 (blue), and S128R1Z4 (red) calculations.
In the top panel we plot $e_{z,y}(k)$ (solid) and $e_{y,z}(k)$
(dotted).  The curves in the bottom panel are $e_{z,x}(k)$ (solid),
$e_{x,z}(k)$ (dotted), and $s$ (dashed).  With this normalization,
curves for different simulations lie nearly on top of each other at
low $k$.  All quantities have been averaged from 50 to 300 orbits.
\label{f:emf}}
\end{figure}

Although it is useful to focus on the vertical wave vectors when
trying to understand properties of large scale fields, an
understanding of the overall power spectrum benefits from an analysis
of the shell integrated quantities.  We plot the time and shell
average EMFs terms in Figure \ref{f:emf} for S32R1Z4 (black) S64R1Z4
(blue), and S128R1Z4 (red).  As in Figure \ref{f:emft}, these
quantities are normalized by the shell integrated power spectra.  Each
normalized term is then time averaged from 50-300 orbits.  Since the
amplitudes of the magnetic energy densities seem to be in statistical
steady states over this period, we presume the left hand sides of
(\ref{eq:fourx}) and (\ref{eq:foury}) are nearly zero.  Therefore the
sum of the terms in each panel must be balanced by numerical
dissipation terms, as discussed in previous work \citep{fp07,shb09}.

In the top panel we plot the terms $e_{z,y}(k)$ (solid) and
$e_{y,z}(k)$ (dotted) which contribute to evolution of $B_x$.  At the
large scales, we find that the $e_{z,y}$ term is more important for
field generation and its normalized amplitude is nearly independent of
resolution. The $e_{y,z}$ term is smaller in amplitude and slightly
negative as large scales.  Even though $e_{y,z}$ tends to oscillates
about zero over an individual dynamo cycle while $e_{z,y}$ usually
remains positive, the amplitude of $e_{z,y}$ is generally larger, so
the dominance of $e_{z,y}$ at large scales is {\it not} simply the
result of time averaging.  As one moves to smaller scales, $e_{y,z}$
rises and eventually dominates the generation of $B_x$.  The
characteristic wavenumber at which the crossing occurs shifts to
higher values as the resolution increases.

The bottom panel shows $e_{z,x}(k)$ (solid), $e_{x,z}(k)$ (dotted) and
$s$ (dashed), the terms which contribute growth in $B_y$.  At large
scales growth of $B_y$ is dominated by the shear term, while both
$e_{z,x}$ and $e_{x,z}$ are of comparable magnitude and negative.  At
small scales, $e_{z,x}$ and $e_{x,z}$ both grow, becoming positive and
dominating over the shear term.  Again, the wavenumber of the
crossover increases with resolution.

Authors have often focused on horizontally average properties of the
flow or (equivalently) the power spectral variation only along
vertical wave vectors \citep[e.g][ which were discussed
above]{fp07,lo08}.  We note that the behavior of $e_{y,z}$ and
$e_{z,y}$ we have described differs significantly from what one would
infer if only vertical wavevectors were considered.  As previously
mentioned, the symmetries of the periodic box force $e_{z,y}(k_z)$ to
be zero, and only $e_{y,z}(k_z)$ contributes.  However, it is clear
from Figure \ref{f:emf} that the vertical EMF and its toroidal
variation is also essential for understanding the mechanism which
sustains turbulence in these simulations.

The question remains as to why the addition of stratification leads to
convergence in the turbulent stresses and energy densities.  One
possibility is that development of local toroidal field is key to
sustaining turbulence in both stratified and unstratified domains.  It
is possible that the strength of toroidal field is entirely set by the
resolution in unstratified domains, while stratified domains offer a
characteristic scale which is independent of resolution, due to the
action of the large scale dynamo.  Indeed, it has already been
demonstrated \citep[e.g.][]{hgb95,sh09} that a global net toroidal
field leads to enhanced turbulent energy densities and stresses, and
leads to convergence in the stress as resolution increses \citep{gua09}.
Furthermore, our simulations show a correlation between the stress and
the strength of the mean toroidal field, both globally in the two
scale height averages (Table \ref{t:sims}) and locally in the
spacetime plots (Figures \ref{f:st128} and \ref{f:st64r4}). This
hypothesis will be addressed further in future research, comparing in
detail the results presented here with those from unstratified runs
both with and without mean fields.

Due to our choice of periodic vertical boundaries, and our use of
simplified thermodynamics, we have largely avoided detailed discussion
of observational implications.  Such questions are generally better
addressed by studies which include more physically realistic vertical
boundary conditions \citep[e.g.][]{ms00} or more realistic
thermodynamics, including the treatment of radiation
\citep[e.g.][]{tur04,hks06}. However, it is worth briefly noting that
our work confirms some important results of earlier studies
\citep[see e.g.][]{bra95,sto96,ms00}.  Figures \ref{f:pscomp} and
\ref{f:st128} show that a significant fraction of the magnetic energy
in these simulations resides in large scale magnetic fields that rise
buoyantly to the low density regions above the disk midplane.
\citet{bp09} argue that the magnetic field structures that power
accretion disk coronae must be associated with characteristic lengths
that are large compared to the typical turbulent eddies. If this were
not the case, the timescales associated with turbulent diffusion would
be smaller than the corresponding buoyant rise time, making it difficult to
transport significant magnetic energy to the coronae.  In other words,
if the corona is a consequence of magnetic field structures that
originate within the turbulent disk via the MRI (or other magnetic
instabilities), but that dissipate above the disk midplane, then these
structures must be of large enough scale to survive the buoyant rise
without being shredded by the turbulence within the disk.  Thus, the
results presented in this paper provide support to the prevailing
paradigm for X-ray emission in accreting systems which involves an
optically thin, hot corona powered by the dissipation of magnetic
fields \citep[e.g.][]{hm93a,fr93}.

\section{Conclusions}
\label{conc}

We have used Athena to examine the effects of stratification on
magnetohydrodynamic turbulence driven by the magnetorotational
instability.  We have shown that stratified simulations converge as
resolution increases, even in domains with zero-net-flux and no
explicit dissipation. This is contrary to our own calculations of
zero-net-flux unstratified domains, which do not converge, confirming
previous results \citep{fp07,gua09,shb09}.  We have also considered
calculations with explicit dissipation, and confirmed previous results
that the maintenance of sustained turbulence is magnetic Prandtl
number dependent.  Stratification appears to extend the range for
which sustained turbulence develops, and may allow sustained
turbulence at slightly lower Prandtl number for a given Reynolds
number. However, the behavior is rather complex with larger variations
and evolution on long timescales (greater than 100 orbits).

At the highest resolutions considered ($64/H$ and $128/H$) the ratio
of total stress to midplane pressure has a mean value of $\alpha \sim
0.01$, but with considerable fluctuation about this mean on long
($\gtrsim 50$ orbit) timescales.  Since real astrophysical systems are
stratified, this somewhat alleviates concerns that magnetorotational
turbulence might be unable to provide the required angular momentum
transport in accretion disks, although values a factor of ten higher
have been inferred in some astrophysical sources \citep{kpl07}.
Similarly, it partially alleviates concerns that explicit dissipation
may be required in global disk simulations at high resolution, as
stratification and net toroidal fields arise naturally in such
calculations.

We have shown that these conclusions do not depend sensitively on the
vertical or radial dimensions of the box.\footnote{Variations in the
  azimuthal length were not considered here.} Domains with radial
extents of one and four scale heights give the same time averaged
values for $\alpha$ and have nearly identical power spectral densities
for the magnetic energy.  Stresses are somewhat more sensitive to
variations in the vertical height of the domain, although this may be
related to our assumptions of vertical periodicity.  Increasing the
vertical extent from four to six scale heights results in only a
slight increase in the time and spatially averaged stresses, as long
as the spatial average is carried out over the same volume (about 10\%
when using the inner two scale heights).

Our results generally reproduce the qualitative features found by
previous authors for stratified systems \citep[e.g.][]{bra95,sto96}.
This includes oscillations with a periods of $\lesssim 10$ orbits in
which the horizontally averaged radial and toroidal fields alternate
sign.  Coupled with buoyancy this leads to a characteristic butterfly
diagram in horizontally averaged space-time plots.  A comparison of
our results with those of \cite{lo08} suggest the mechanisms for
generating the large scale field oscillations in the stratified and
unstratified domains may be related.

\acknowledgements{We thank C.-K. Chan for useful conversations.  We
  also thank N. Lemaster and J. Simon for providing their Fourier
  analysis codes which were used as the starting point for our code,
  and for comparison purposes, respectively.  SWD is supported by NASA
  grant number PF6-70045, awarded by the Chandra X-ray Center, which
  is operated by the Smithsonian Astrophysical Observatory for NASA
  under contract NAS8-03060.  MEP and SWD acknowledge support by the
  Institute for Advanced Study, and JMS acknowledges support from DOE
  grant number DE-FG52-06NA26217 and NASA grant number NNG06GJ17G.
  Computations were performed on facilities provided by the Princeton
  Institute for Computational Science and Engineering and the
  Institute for Advanced Study's aurora cluster.}

\end{document}